\documentclass[lettersize,journal]{IEEEtran}
\usepackage{amsmath,amsfonts}
\usepackage{algorithmic}
\usepackage{array}
\usepackage[caption=false,font=normalsize,labelfont=sf,textfont=sf]{subfig}
\usepackage{textcomp}
\usepackage{stfloats}
\usepackage{url}
\usepackage{verbatim}
\usepackage{graphicx}
\usepackage{cite}
\usepackage{amsmath,amssymb,amsfonts}
\usepackage{algorithmic}
\usepackage{textcomp}
\usepackage{multirow}
\usepackage{svg}
\usepackage[linesnumbered,ruled,vlined]{algorithm2e}

\usepackage{xcolor}

\usepackage{xspace}
\newcommand{\tools}[1]{\textsc{#1}}
\newcommand{\kissat}{\tools{Kis\-sat}\xspace}
\newcommand{\MAB}{\tools{Kis\-sat-MAB}\xspace}
\newcommand{\dpCEC}{\tools{hybridCEC}\xspace}
\newcommand{\parCEC}{\tools{paraHCEC}\xspace}
\newcommand{\kcbox}{\tools{KCBox}\xspace}
\newcommand{\minisat}{\tools{MiniSat}\xspace}

\def\BibTeX{{\rm B\kern-.05em{\sc i\kern-.025em b}\kern-.08em
    T\kern-.1667em\lower.7ex\hbox{E}\kern-.125emX}}
\usepackage{balance}
\begin{document}
\title{Datapath Combinational Equivalence Checking With Hybrid Sweeping Engines and Parallelization}
\author{\IEEEauthorblockN{Zhihan Chen$^{1,2}$, \;\; Xindi Zhang$^{1,2}$, \;\; Yuhang Qian$^{1,2}$, \;\; Shaowei Cai$^{1,2,*}$} \\
\IEEEauthorblockA{
\textit{1.School of Computer Science and Technology, University of Chinese Academy of Sciences, Beijing, China} \\
\textit{2.Key Laboratory of System Software (Chinese Academy of Sciences) and State Key Laboratory of Computer Science, Institute of Software, Chinese Academy of Sciences, Beijing, China}\\
\{chenzh, zhangxd, qianyh\}@ios.ac.cn; caisw@ios.ac.cn}
\thanks{Manuscript created October, 2020; This work was developed by the IEEE Publication Technology Department. This work is distributed under the \LaTeX \ Project Public License (LPPL) ( http://www.latex-project.org/ ) version 1.3. A copy of the LPPL, version 1.3, is included in the base \LaTeX \ documentation of all distributions of \LaTeX \ released 2003/12/01 or later. The opinions expressed here are entirely that of the author. No warranty is expressed or implied. User assumes all risk.}}

\markboth{Journal of \LaTeX\ Class Files,~Vol.~18, No.~9, September~2020}%
{How to Use the IEEEtran \LaTeX \ Templates}

\maketitle

\begin{abstract}
In the application of IC design for microprocessors, there are often demands for optimizing the implementation of datapath circuits, on which various arithmetic operations are performed. Combinational equivalence checking (CEC) plays an essential role in ensuring the correctness of design optimization. The most prevalent CEC algorithms are based on SAT sweeping, which utilizes SAT to prove the equivalence of the internal node pairs in topological order, and the equivalent nodes are merged. Datapath circuits usually contain equivalent pairs for which the transitive fan-in cones are small but have a high XOR chain density, and proving such node pairs is very difficult for SAT solvers. An exact probability-based simulation (EPS) is suitable for verifying such pairs, while this method is not suitable for pairs with many primary inputs due to the memory cost. We first reduce the memory cost of EPS and integrate it to improve the SAT sweeping method. Considering the complementary abilities of SAT and EPS, we design an engine selection heuristic to dynamically choose SAT or EPS in the sweeping process, according to XOR chain density. Our method is further improved by reducing unnecessary engine calls by detecting regularity. Furthermore, we parallelized the SAT and EPS engines of \dpCEC, leading to the parallel CEC prover. Experiments on a benchmark suite from industrial datapath circuits show that our method is much faster than the state-of-the-art CEC tool namely ABC `\&cec' on nearly all instances, and is more than 100$\times$ faster on 30\% of the instances, 1000$\times$ faster on 12\% of the instances. In addition, the 64 threads version of our method achieved 77x speedup.
\end{abstract}


\section{Introduction}

Combinational Equivalence Checking (CEC) is the problem to formally prove whether two design specifications are functionally equivalent, which is one of the most essential techniques in Electronic Design Automation (EDA) and digital IC design. It has a wide range of applications, such as functional equivalent logic removal~\cite{amaru2020sat}, sequential equivalence checking~\cite{mishchenko2008scalable},
circuit-based method for symmetries detection~\cite{zhang2006symmetry},
engineering change orders~\cite{jarratt2011engineering}, 
among others.

Datapath circuits, consisting of many fundamental arithmetic units such as multipliers, adders, and multiplexers, are commonly shown in computationally intensive applications like micro-processor design~\cite{nowka1998circuit}.
To achieve better PPA~(Performance, Power, and Area), datapath circuits are optimized by logic synthesis or structural rewriting, and CEC is an essential tool for ensuring the correctness of the optimizations.

State-of-the-art CEC algorithms, e.g., the ABC\cite{mishchenko2007abc} `\&cec' command, are based on SAT sweeping~\cite{zhu2006sat,mishchenko2006improvements}. For two circuits to be verified, a miter circuit is created by pairwise connecting the PIs and POs, and the two circuits are equivalent iff the miter's output is always 0. This can be checked by calling a SAT solver on the formula of the miter circuit. Before performing SAT sweeping, logic simulation is used to generate a series of internal node pairs that are potentially equivalent.
In SAT sweeping, the equivalence of each internal node pair is checked with SAT solvers, and the equivalent nodes are merged once verified.

When sweeping the datapath circuits, there are usually internal node pairs for which the transitive fan-in cone of the corresponding miter circuit has a high density of XOR chains. Verifying such nodes is a challenge for modern SAT solvers, even for small cones. There is a method tailored for sweeping such high XOR chain density circuits, known as exact probabilistic-based simulation (EPS) method~\cite{wu2006potential}, which works particularly well for such circuits of small size (up to 24 primary inputs). For example, modern SAT solvers cannot prove the equivalence of the middle primary output of two 16-bit multipliers in 12 hours, while EPS can finish the verification in 4 minutes. However, EPS needs $O(2^{N})$ memory to save the probability signal for each node, where $N$ is the number of primary inputs (PI). This limits its usage in applications, and thus we rarely see evidence of its application as a standalone CEC tool.

In this work, we incorporate EPS into SAT Sweeping to make up for the weakness of the SAT solver on circuits of a high XOR chain density. Additionally, to improve the scalability of EPS, we propose to divide the excessively long assignments into multiple groups of shorter signals.

In our method, we combine SAT and EPS for sweeping. This is motivated by the observation that SAT and EPS exhibit very clearly complementary performance, depending on the density of XOR chains in the circuits ---  EPS performs essentially better than SAT solvers on those with high XOR chain density, while SAT solvers are far better than EPS on those with low XOR chain density.
Therefore, we design a heuristic to dynamically select the engine (SAT or EPS) for sweeping, according to the XOR chain density of the corresponding cones of the nodes.

Besides, we design an identical structure detection~(ISD) technique to discover sub-modules with the same implementation and thus reduce redundant sweepings.  This leads to further improvement, as there are usually many sub-modules with the same implementation in structural and topological levels (regularity) for datapath circuits~\cite{chowdhary1999extraction}. Eventually, an efficient CEC prover \dpCEC was formed.

However, it is still challenging for current sequential CEC provers to meet the demand for expanding scales and difficulties from industrial and academic datapath circuits. Parallelism is a viable and vital strategy.
There have been many parallelism CEC provers.
Chatterjee et. al. proposed EQUIPE~\cite{chatterjee2010equipe}, which leverages the massive parallelism of modern general-purpose graphic processing units, and reduces the number of the engine (SAT and BDD) calls. However, with 14 GPU-cores and 4 CPU-cores, EQUIPE can only improve by three times compared to the sequential ABC tool. Recently, Possani1 et. al.~\cite{possani2019parallel} proposed three complementary ways for enabling parallelism in CEC, which are graph partitioning, main miter partitioning and internal miter partitioning. The runtime is significantly reduced for the large-scale circuits compared to the sequential ABC tool.
But to our knowledge, no previous parallel provers are aiming at datapath circuits. 
The differences between datapath circuits and general circuits lie in that:
Proving the internal pairs usually takes a long time for datapath circuits, while for general circuits, the number of pairs is large, but the proof is fast.
General-purpose solvers focus on parallelism on the sweeping framework, neglecting the parallelism of internal engines, which is usually the bottleneck of the whole process.

On top of \dpCEC, we parallelized two solving engines SAT and EPS, leading to a parallel CEC prover \parCEC. Specifically, for SAT, we replaced the sequential SAT solver in \dpCEC with an efficient parallel solver PRS~\cite{chen2023prs}. For EPS, by enumerating the values of any $k$ PIs, the overall task can naturally be divided into $2^k$ uniform subtasks, thus achieving efficient parallelization.

We evaluate our prover, \dpCEC and \parCEC, on an industrial benchmark of 50 datapath circuits.
We experimentally compare our prover with state-of-the-art CEC tools, namely ABC `\&cec', as well as the pure-SAT/BDD methods.
Experiments show that \dpCEC is almost always faster than competitors. Especially, our prover is 100$\times$ faster than ABC `\&cec' on 30\% of the instances, and 1000$\times$ faster on about 12\% of the instances. It proves 23 more instances than ABC `\&cec'. When compared with the pure-SAT/BDD method (checking the miter circuit by calling a SAT/BDD solver just one time), our prover also shows similar superiority.

In addition, the parallel implementation significantly enhanced the performance of \dpCEC. Especially, the speedup of \parCEC with 2 (4, 8, 16, 32, 64) threads is 3.3 (6.0, 11.4, 21.3, 43.3, 77.4) according to the PAR2, which verifies the effectiveness and scalability of \parCEC.

This paper extends our previous work~\cite{chen2023integrating}, which proposes the sequential hybrid prover \dpCEC. The new contributions of this paper include a parallel hybrid CEC prover \parCEC, extensive related experiments, and detailed analysis of the prover. 

\noindent
\textbf{Paper Organization:}
Section~\ref{sec:pre} introduces the preliminary and background. Section~\ref{sec:contri} gives the framework of \dpCEC and the contributions. Section~\ref{sec:exp} shows the experimental results and evaluates the proposed techniques of \dpCEC. Section~\ref{sec:parahcec} introduces the parallelization strategies of \parCEC. Section~\ref{sec:paraexp} presents the experimental results and detailed analysis of \parCEC. Finally, Section~\ref{sec:concl} concludes this work.

\section{Background and Related Works\label{sec:pre}}

This section introduces fundamental definitions, some important techniques, and  related works.

\subsection{Preliminaries for Circuits and CEC}

A \emph{Boolean network} is a directed acyclic graph (DAG) $G(V, E)$ with a set $V$ of Boolean variables (node) and a set $E\subseteq V \times V$ of edges that determine the topology of the variables. 
A combinational gate-level \emph{netlist} is a Boolean network, in which wires are associated with Boolean variables in $V$ and gates are associated with Boolean functions represented by several edges in $E$. 
In this paper, the terms `network' and `netlist' are used to denote a circuit.
There are primary inputs (PI) and primary outputs (PO) for each Boolean network. Given an edge $e = (u, v)$, $u$ is the \emph{fan-in} of $v$ while $v$ is the \emph{fan-out} of $u$. A \emph{transitive fan-in} (TFI) \emph{cone} of a node $v \in V$ is the collection of nodes that terminated at $v$ based on some set of edges in $E$, which can be seen as a sub-module of the circuit.

We focus on the gate-level netlist in the format of \emph{And-Invert-Graphs} (AIG), where each node is a two-input AND gate and each input of the gate may be associated with an inverter.
AIG is a generic and tidy format that can easily represent any arbitrary Boolean network.
Combinational Equivalence Checking~(CEC) is a problem for checking whether two given combinational circuits are functionally equivalent. In this paper, we focus on checking the equivalence of two datapath circuits, which consist of many arithmetic units such as multipliers, adders, multiplexers, among others. Due to the large proportion of AND-XOR terms in the design, CEC on this type of circuit is usually very hard~\cite{lv2012formal}.

\begin{figure}[t]
\centerline{\includegraphics[width=0.3\textwidth]{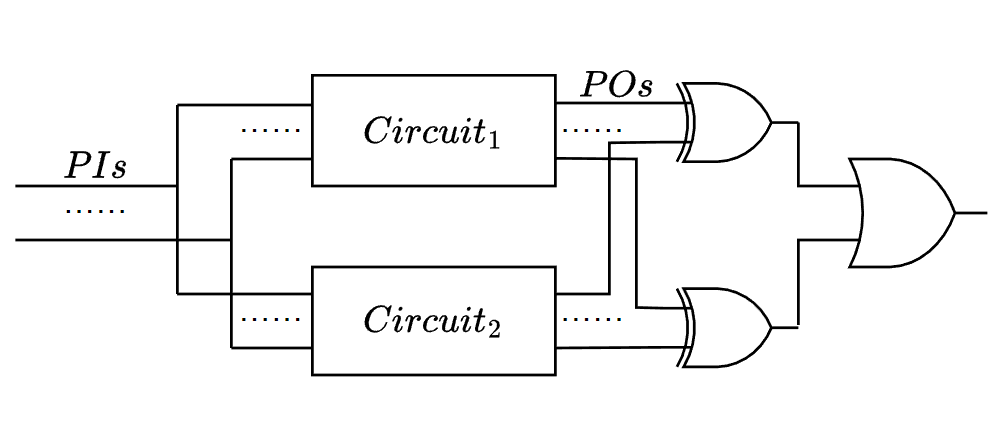}}
\caption{The structure of miter circuits. \label{fig:miter}}
\label{fig}
\end{figure}

Given two circuits with the same number of PIs and POs, a \emph{miter} circuit~\cite{brand1993verification} is composed of the two circuits to be compared by pair-wise connecting the PIs and POs. As shown in Figure~\ref{fig:miter}, the matched PIs are connected directly while the matched POs are connected by XOR gates, and an OR gate connects the output of newly inserted XOR gates. For a given miter, the original two circuits are equivalent if the output of the OR gate at the right of the miter is always a constant 0.

\subsection{Reasoning Tools}

Given a set of Boolean \emph{variable}s $V=\{x_1,x_2,...,x_n\}$, a \emph{literal} is either the positive or negative of a Boolean variable. A \emph{clause} $C = \bigvee _i x_i$ is a disjunction of literals. A \emph{Conjunctive Normal Form} (CNF) formula $F = \bigwedge _i C_i$ is the conjunction of clauses. An \emph{assignment} is a mapping $\alpha: V \rightarrow \{0, 1\}$ that assigns values to the variables in $V$.
The Boolean Satisfiability problem~(SAT) is a decision procedure for deciding whether there is an assignment that satisfies the given CNF formula. SAT is the most popular reasoning tool in current CEC tools.

A circuit in the AIG format can be naturally encoded into CNF, where the AIG variable can be mapped into the Boolean network and the gates can be encoded into several clauses by the Tseitin transformation~\cite{tseitin1983complexity}. For a miter circuit to be verified, the assertion that `output should be a constant 0' is encoded into a unit clause of the CNF formula.

The Conflict-Driven Clause Learning~(CDCL) framework is widely used in modern SAT solvers~\cite{marques1999grasp}. The incremental SAT solving technique is designed for continuously solving a series of similar problems~\cite{een2003temporal}.
The most prevalent and powerful CDCL solvers are \kissat~\cite{fleury2020cadical} and their derived versions.

Binary Decision Diagram~(BDD) is another primary inference tool in the proof procedures of CEC, and BDD tools~\cite{lai2021power} accept CNF as their input file as well. BDD is efficient for small-scale circuits and was prevalent before the year 2000; however, it is difficult to verify large arithmetic circuits due to memory limitations. SAT solvers gradually become the mainstream method because of their scalability for large instances. There are works that incorporate both SAT and BDD solvers~\cite{kuehlmann2002robust}. 

\subsection{Sweeping-based CEC Flow}
\begin{figure}[t]
\centerline{\includegraphics[width=0.35\textwidth]{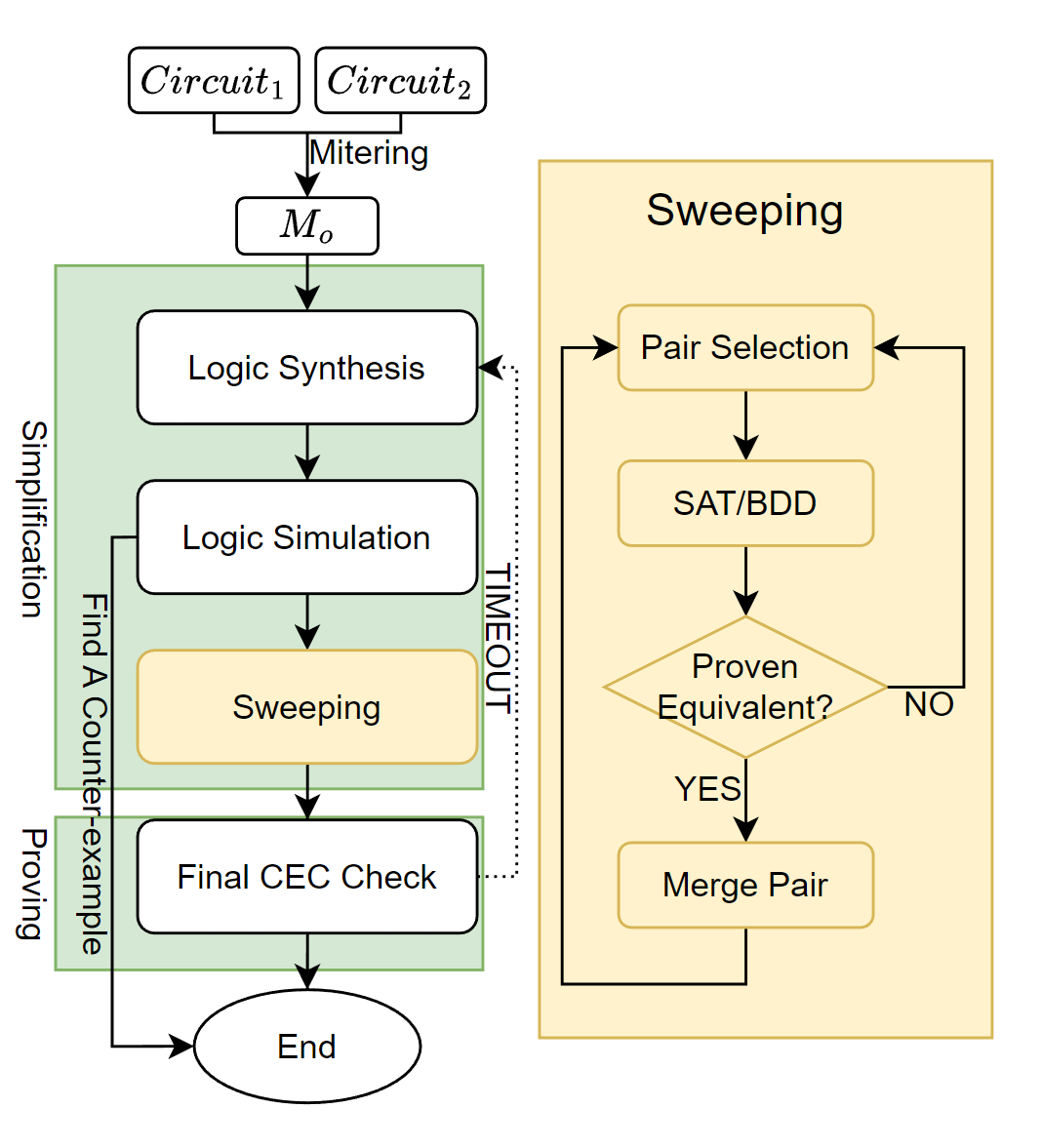}}
\caption{Framework for a Typical Sweeping Based CEC Algorithm.}
\label{fig:framework1}
\end{figure}

The most prevalent method of current CEC engines is the sweeping-based method, and a typical flow of this method is shown in Figure~\ref{fig:framework1}. 
Given two circuits to be verified, the first step is to construct a miter $M_o$.
Then the main CEC engine will alternative between `$M_o$ proving stage' to check the equivalence of the given two circuits and `$M_o$ simplification stage'.

\begin{figure}[t]
\centerline{\includegraphics[width=0.3\textwidth]{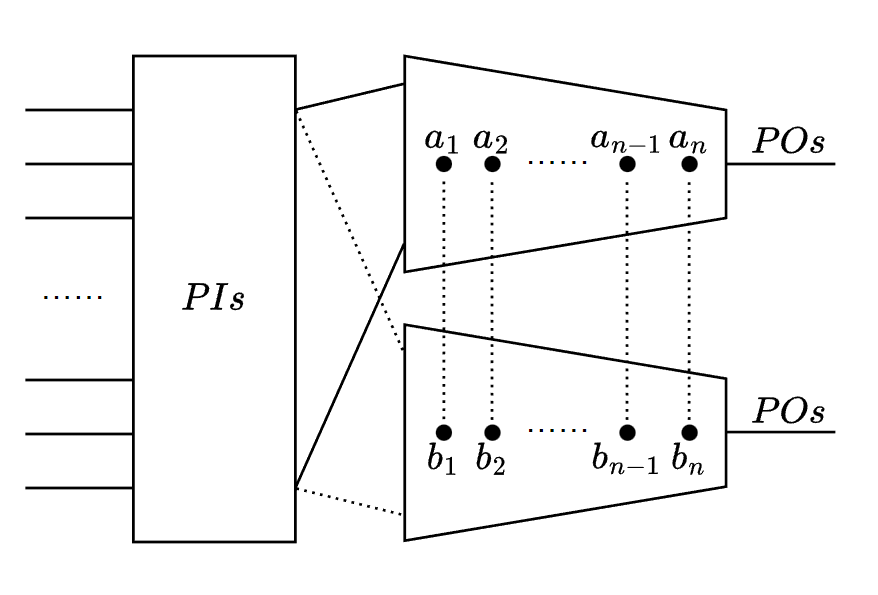}}
\caption{A Schematic Diagram for Sweeping.}
\label{fig:sweeping}
\end{figure}

In the miter simplification stage, $M_o$ is simplified by logic synthesis~\cite{bjesse2004dag}, structural hashing~\cite{kuehlmann2002robust}, functionally reduced AIGs~\cite{mishchenko2005fraigs}, BDD sweeping~\cite{kuehlmann2002robust}, or SAT sweeping~\cite{kuehlmann2004dynamic}. 
Sweeping is the most popular method, and the schematic diagram is shown in Figure\ref{fig:sweeping}. It first detects some potential-equivalent internal node pairs $(a_1,b_1), ... (a_n,b_n)$ by \emph{logic simulation}, then the equivalence of each node pair $(a_i, b_i)$ are checked with a reasoning tool (SAT/BDD) following a topological order. Once a node pair $(a_i,b_i)$ is proven to be equivalent, the two nodes can be merged into a single node, as a result, the original miter $M_o$ is reduced and then it becomes easier to prove.

Logic simulation, a pre-procedure of sweeping, is used to group nodes that
have the potential to be equivalent into classes. It applies random values to the primary inputs (PIs) of the circuit and propagates these values toward the primary outputs (POs). From the value associated with each node, we can learn the logical behavior of internal nodes, and quickly rule out non-equivalent node pairs.
Sometimes, the logic simulation could find counter-examples of two non-equivalent circuits directly.
Typically, the value assigned to PIs can be generated heuristically based on information from the SAT engines during the sweeping process~\cite{mishchenko2006improvements}.

\subsection{Exact Probability-based Simulation \label{sec:eps}}

Different from the logic simulation, which applies and propagates Boolean values, probability-based simulation methods~\cite{agrawal1996characteristic, wu2006potential} applies numeric probability signal $p\in[0,1]$ for each 
PI, and fetch the probability signal at the PO by probability calculations. In detail, for an inverter, if the probability for the input is $p$, then the probability of the output is $1-p$;  for a two-input AND gate, if the probabilities of the two inputs are $p_1$ and $p_2$, then the output probability is $p_1\times p_2$.

The output probabilities can be viewed as characteristics or properties of the logical function of a network~\cite{agrawal1996characteristic}, which is used to identify the logical function of a network. That means the output probabilities of two logical equivalent networks will be identical.
However, a phenomenon called \emph{aliasing} may occur, where networks with different logical functions share the same output probabilities.

In \cite{wu2006potential}, Shih-Chieh Wu et al. proposed a method that could generate aliasing-free probability assignments which can ensure the absence of `aliasing'. Two networks are equivalent iff their output probabilities are the same under the same aliasing-free assignments.

The exact probability-based simulation~(EPS) method proposed in~\cite{wu2006potential} is described as follows.


For a miter with $N$ PIs, the probability for the $i$-th PI should be assigned to $1/\theta_i$ according to Formula~\ref{fml:epec}, where $\theta_1$ is assigned to 3 to minimize memory usage.
\begin{equation}
    \label{fml:epec}
    \theta_{i+1} \leftarrow \left(\theta_i - 1\right)^2 + 1, i \in [1, N-1];
\end{equation}

A more efficient calculation method is applied in practice~\cite{wu2006potential}, which uses fast bitwise operations instead of using heavy float multiplication. After generating a set of aliasing-free probabilities for PIs, all the fractions of the PI probabilities are reduced to a common denominator, then we could focus on the numerator, which are long integers that are proper to perform bitwise operations. Then the values are propagated using bitwise-AND for AND gates and bitwise-NOT for inverters. 

However, the disadvantage of EPS is that the length of the assignments exponentially grows according to the number of PIs. 
In detail, for a circuit with $N$ PIs, the length of the numerator is $2^N$ bits with the bitwise optimized method.
Thus, this approach is not suitable for circuits with large $N$ due to the large memory overhead.

\section{Sequential CEC Hybrid Method\label{sec:contri}}

This section presents our CEC method. We start with a motivating example for our method.  Then, we introduce the main CEC framework of our method, followed by the three main ideas, including improved exact simulation, engine selection heuristic, and identical structure detection.

\subsection{Motivating Example\label{sec:motivation}}

Checking the equivalence of datapath circuits usually appears in the application of the optimization of the microprocessor. For the elaboration of our motivation, we use a miter circuit named `ec\_h1' in this subsection, which is an industrial instance from designing long bit-wise arithmetic circuits. 

ABC `\&cec' is not capable of solving this instance. By applying $2^{20}$ rounds logic simulation randomly (less than 0.1 seconds), 113 potential-equivalent node pairs are selected. Finally, after removing redundant regularity pairs whose cone has the same structure as a previous pair's, 48 node pairs are chosen.

In a typical SAT sweeping flow, the miter circuits of the 48 node pairs are translated into CNFs and checked by a SAT solver. We use a state-of-the-art SAT solver \MAB~\cite{cherif2021kissat} for testing. For the detailed experiment settings, please refer to Section~\ref{sec:exp}.

According to the results in Table~\ref{tbl:pre}, we noticed that there are 10 node pairs that are too hard for SAT solvers (we also tried MiniSAT, but only with worse results), while they can be solved easily by EPS. 
On the other hand, there are 16 node pairs that cannot be proven by EPS, which can be solved quickly by SAT solvers. From the table, we noticed that EPS has no ability for proving the equivalence of circuits whose PI number is larger than 36, mainly due to the huge memory cost.

From the results, we learn that SAT and EPS are complementary on different circuits. Further observations show that the performance difference is mainly related to the density of XOR chains. This motivates us to design a hybrid Sweeping CEC tool that uses both SAT and EPS.

\begin{table}[t]
\caption{Runtime comparison between SAT and EPS for verifying the internal pairs of miter `ec\_h1'. Pairs that are too easy (less than 1 second) for both SAT and EPS engines are not reported. We report the topologic index (ID), the number of gates (Gates), the number of PIs of the TFI cone (PIs), and the runtime for SAT and EPS. `TO' denotes timeout of 1 hour.
\label{tbl:pre}}
\begin{center}
\setlength{\tabcolsep}{17pt}
\begin{tabular}{|c|c|c|c|c|}
\hline
\multirow{2}*{ID} & \multirow{2}*{Gates} & \multirow{2}*{PIs} & \multicolumn{2}{|c|}{ Reasoning Tools} \\
\cline{4-5} 
 & &  & SAT & EPS \\
\hline
17 & 268 & 32 & \textbf{$<$0.01} & 9.67\\
18 & 353 & 18 & 1.75 & \textbf{$<$0.01}\\
19 & 360 & 36 & \textbf{$<$0.01} & TO\\
20 & 452 & 20 & 5.47 & \textbf{$<$0.01}\\
21 & 556 & 22 & 29.40 & \textbf{0.02}\\
22 & 468 & 40 & \textbf{$<$0.01} & TO\\
23 & 548 & 44 & \textbf{$<$0.01} & TO\\
24 & 678 & 24 & 137.32 & \textbf{0.08}\\
25 & 658 & 48 & \textbf{$<$0.01} & TO\\
26 & 807 & 26 & 903.51 & \textbf{0.65}\\
27 & 768 & 52 & \textbf{$<$0.01} & TO\\
28 & 1097 & 30 & TO & \textbf{13.30}\\
29 & 950 & 28 & TO & \textbf{2.94}\\
30 & 1423 & 32 & TO & \textbf{66.41}\\
31 & 1310 & 64 & \textbf{$<$0.01} & TO\\
32 & 1022 & 60 & \textbf{$<$0.01} & TO\\
33 & 896 & 56 & \textbf{$<$0.01} & TO\\
34 & 1259 & 32 & TO & \textbf{59.64}\\
35 & 2018 & 32 & TO & \textbf{90.56}\\
36 & 1734 & 32 & TO & \textbf{79.45}\\
37 & 1580 & 32 & TO & \textbf{73.27}\\
38 & 1832 & 64 & \textbf{$<$0.01} & TO\\
39 & 1580 & 64 & \textbf{$<$0.01} & TO\\
40 & 1446 & 64 & \textbf{$<$0.01} & TO\\
41 & 1168 & 64 & \textbf{$<$0.01} & TO\\
42 & 2262 & 32 & TO & \textbf{132.24}\\
43 & 2144 & 32 & TO & \textbf{122.90}\\
44 & 1879 & 32 & TO & \textbf{84.97}\\
45 & 2046 & 64 & \textbf{$<$0.01} & TO\\
46 & 1942 & 64 & \textbf{$<$0.01} & TO\\
47 & 1708 & 64 & \textbf{$<$0.01} & TO\\
48 & 4280 & 96 & \textbf{$<$0.01} & TO\\
\hline

\hline
\end{tabular}
\label{tab1}
\end{center}
\end{table}

\subsection{Main Framework of Our Method}

\begin{figure}[t]
\centering
\centerline{\includegraphics[width=0.25\textwidth]{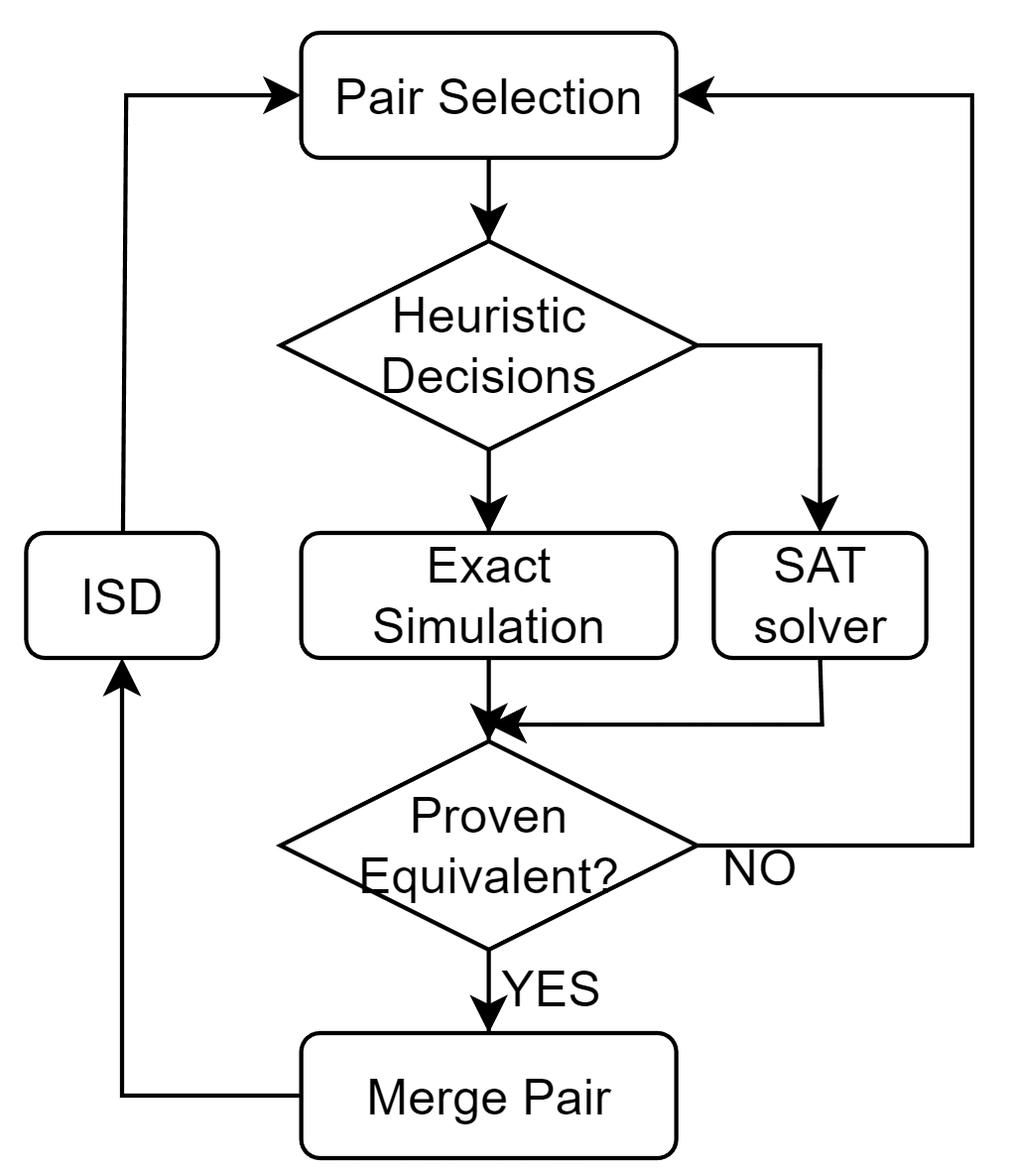}}
\caption{Main Framework for Our Sequential CEC Algorithm.}
\label{fig:framework2}
\end{figure}

The framework for our algorithm, \dpCEC, is similar to the typical SAT sweeping based method presented in Figure~\ref{fig:framework1}, and the primary difference is the sweeping process, as depicted in Figure~\ref{fig:framework2}.

During the SAT Sweeping process, we propose a heuristic to select a reasoning tool (either SAT or exact simulation) for checking the equivalence of two potential-equivalent internal points. Furthermore, when a node pair is successfully merged, we further check the remaining pairs following the topological order; 
if there is a pair ($a_i$, $b_i$), which is in the same implementation as the newly merged equivalent pair, 
then $a_i$ and $b_i$ can be merged immediately. This technique is called Identical Structure Detection (ISD). As a result, an efficient CEC algorithm is developed, which is called \dpCEC, as it combines SAT and exact simulation.


\subsection{Improved Exact Simulation}
\begin{figure}[t]
\centerline{\includegraphics[width=0.43\textwidth]{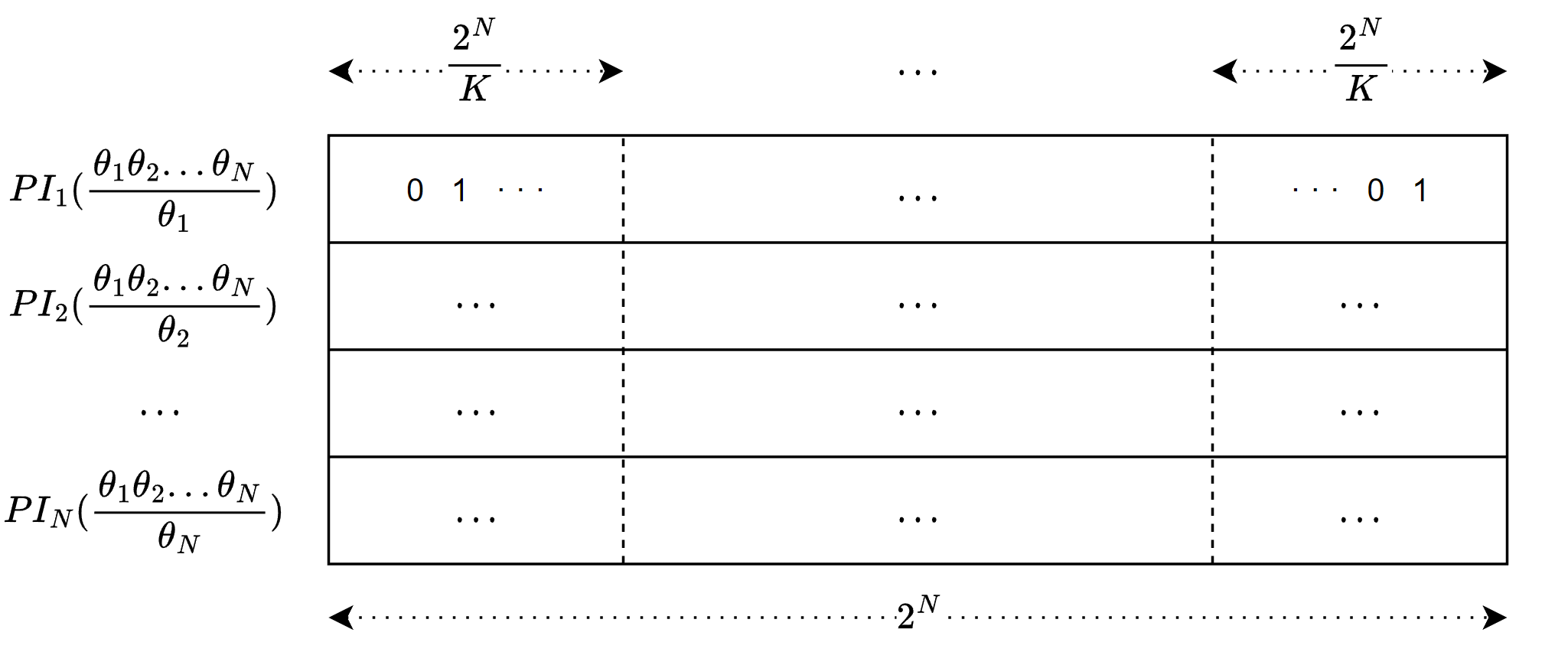}}
\caption{Truncate the PI values into $K$ blocks of small values.}
\label{fig:split}
\end{figure}

As stated in Section~\ref{sec:eps}, the  EPS algorithm in \cite{wu2006potential} assigns $1/\theta_i$ for the $i$-th PI. For a circuit with $N$ PIs, the value $i$-th becomes a large integer $\theta_1\theta_2...\theta_N/\theta_i$ after reducing to a same denominator and eliminating the denominator.

The binary representation of the assigned values for each PI, as shown in Figure~\ref{fig:split}, is highly structured, with each column in the diagram denoting a Boolean assignment pattern for PIs. In total, there are $2^N$ possible assignment patterns, i.e., to promise aliasing-free, the original EPS methods assign an integer with at least $2^N$ bits wide for each node.

When the number of PIs $N$ is large, performing a simulation with the full integer for each PI results in huge memory usage. Therefore, in our approach, we divide the long integer into $K$ groups of smaller integers and perform simulations one group at a time, thus reducing memory consumption. The splitting method is shown in Figure~\ref{fig:split}.

The pseudocode for the improved EPS is shown in Algorithm~\ref{alg:epcec}. In the beginning, the algorithm sorted the gates into a topological order (line 1). Then the length of each group ($l$) and the rounds of simulation needed to perform ($R$) are calculated (lines 2--3). The number of PIs should not exceed the $bits\_limit$. Then, for each group, the short integers are applied to the PIs (lines 5-6), and the values are propagated according to the assignment (lines 7--8).
Meanwhile, SIMD can be used to accelerate propagation. 
Counter-example may be found in the simulation process (lines 9--10).
Once all rounds are completed, all possible patterns of the PIs have been tested, which successfully verified the miter (line 11).

With this truncation method, the memory cost of EPS is limited to a reasonable range. The improved EPS method could check circuits with 32 PIs in 4 minutes, while circuits with 40 PIs would take approximately half day.

\begin{algorithm}[t]
\caption{Improved Exact Simulation}\label{alg:epcec}
\KwIn{Miter network $M$, $PI=\{i_0,i_1,...,i_k\}$, $PO = o$}
\KwOut{Equivalent/ Non-equivalent}

    $gate\_list\gets topological\_sort(M)$

    $l \gets min\{|PI|, bits\_limit\}$

    $R\gets 2^{|PI|-l}$

    \For{$i$ in 1, ..., R} {
        \For{each $p_j$ in $PI$} {

            $V_{p_j} \gets construct\_initial\_value(j, i)$

        }
        \For{each gate $G_j$ in $gate\_list$} {
        
            $propaget\_value(G_j)$
        
        }
        \If{$V_{o}$ is not 0} {
        
            \Return Non-equivalent
    
        }
 
    }
    
    \Return Equivalent
    
\end{algorithm}

\subsection{Selection Heuristic\label{sec:select}}

During the SAT Sweeping process, SAT and EPS have very different performances on circuits with different structures. 
By analyzing the internal design of the instances (including `ec\_h1' mentioned in Section~\ref{sec:motivation}), we noticed that, for those node pairs that are difficult for SAT solvers to prove, the miter circuit of their TFI cones all have a large proportion of  XOR chains.

Therefore, for a miter circuit $M$ to be verified, we design a heuristic  to determine whether to use SAT solvers or EPS, according to the XOR chain density.
We propose a function, denoted as $score_{XOR}$, to measure the XOR chain density, which is defined below.

\begin{equation}
    \label{fml:xor-score}
    score_{XOR}(M) = log_2(\sum_{b \in BS} 2^{|b|}) / N
\end{equation}

$score_{XOR}$  is designed with the following intuition:
The runtime of EPS is exponentially increasing with respect to the number of PIs, which can be estimated as $2^N$.
CDCL has poor performance on unsatisfiable instances with a large proportion of XOR chains, because the unsatisfiable certificates of an XOR chain require an exponential number of clauses in the worst case~\cite{dudek2017hard, gwynne2014sat}. The runtime to refute an XOR block with $b$ XOR gates can be estimated as $2^b$, and the runtime for a circuit with more than one XOR chain can be estimated as $\sum_{b \in BS} 2^{|b|}$.
Though SAT is an NP-hard problem, practical CDCL solvers are more efficient than the theoretical estimates thanks to various optimization in decades years. Therefore, a coefficient $\rho$ is introduced to balance the gap. In this paper, $\rho$ is set to 0.15 according to experiments.

With the $score_{XOR}$ function,  the selection heuristic is shown in Algorithm~\ref{alg:selection}. For a given node pair $(a_i, b_i)$ to be checked, we first fetch the TFI cones of the two nodes and construct a miter $M$ of the cones (line 1). Then, we recognize the XOR gates in $M$ and group the XOR gates into XOR blocks $BS$ according to the connection relationship (line 2). The scoring function is calculated by visiting each XOR block in $BS$ (lines 3--6). Finally, the heuristic prefers to pick EPS when the XOR chain density is larger than a parameter $\rho$; otherwise, SAT is picked. 







\begin{algorithm}[t]
\caption{Selection Heuristic}\label{alg:selection}

\KwIn{Node pair $(a_i, b_i)$}
\KwOut{Reasoning tool (SAT / EPS)}
\SetAlgoLined
\setcounter{AlgoLine}{0}
    
    $M\gets miter\_TFI\_cones(a_i, b_i)$
    
    $BS\gets find\_all\_XOR\_blocks(M)$

    $score_{XOR}\gets 0$
    
    \For {$b$ in $BS$} {
        $score_{XOR} \gets score_{XOR} + 2^{|b|}$    
    }

    $score_{XOR} \gets \frac{log_2(score_{XOR})}{|PI|}$

    \If {$score_{XOR} > \rho$} {
    
        \Return EPS
    
    }
    \Else {
        
        \Return SAT
    
    }

\end{algorithm}

\subsection{Identical Structure Detection}
In the sweeping process, numerous internal potential-equivalent nodes are selected by logic simulation, and most of them are truly equivalent nodes. 
The arithmetic units of datapath circuits are usually typical ones, such as multipliers and adders. Consequently, many arithmetic units share the same logic function and implementation, i.e., with regularity~\cite{chowdhary1999extraction}. 

For each potential-equivalent pair, if a previous pair, which has the same structural and topological implementation (regularity), has been proven to be equivalent, then there is no need to recheck the equivalence with the time-consuming reasoning tools again.

\begin{figure}[t]
\centerline{\includegraphics[width=0.22\textwidth]{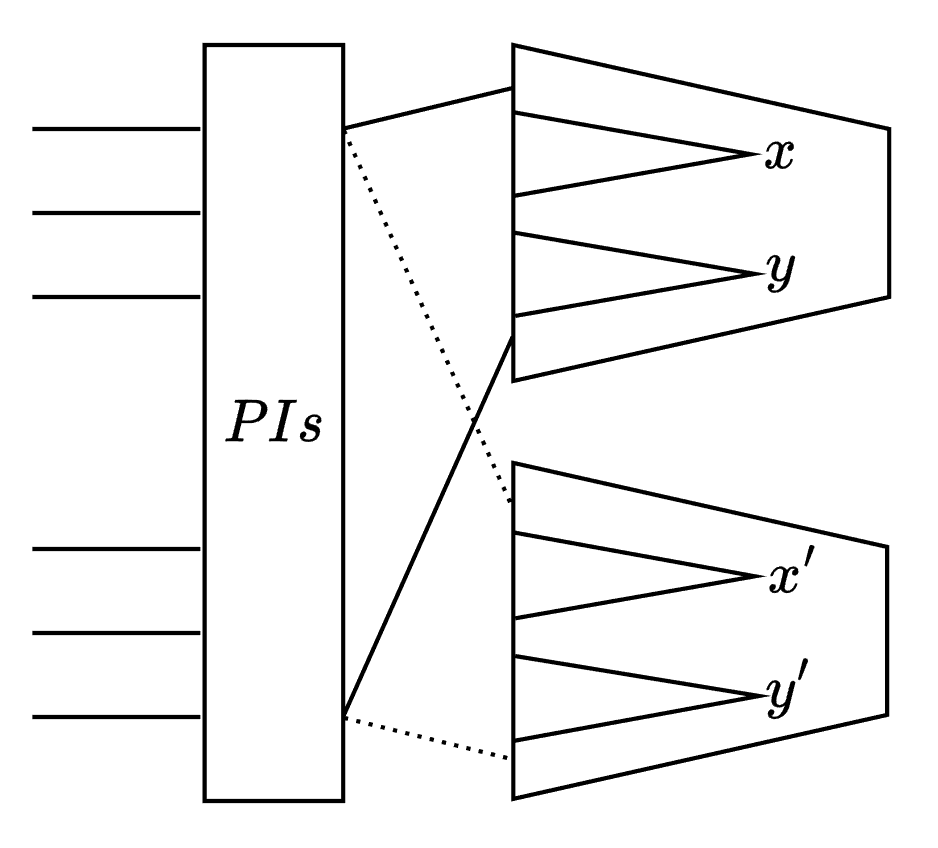}}
\caption{Schematic Diagram for Identical Structure Detection.}
\label{fig:asd}
\end{figure}

The schematic diagram is shown in Figure~\ref{fig:asd}, let node pair $(x, x')$ are functional equivalent, which is proved by a single call of a reasoning tool. If nodes $x$, $y$ are two units with the same implementation, and nodes $x'$, $y'$ are two units with the same implementation as well, then the node pair $(y, y')$ are the functional equivalent nodes.

The specific method for regularity checking for the TFI cones of two nodes, $x, y$, is a greedy algorithm. Initially, we maintain a queue with only one node pair $(x, y)$. If the queue is not empty, we fetch a node pair $(a, b)$ from the queue, and the inputs of the AND gates whose outputs are $a$ and $b$ are compared. Let the inputs of $a$ is $a_1$ and $a_2$, the inputs of $b$ is $b_1$ and $b_2$. We check whether the polarity of $a_1$ and $b_1$ are matched, and the same for $a_2$ and $b_2$. If they are both matched, we push the two pairs ($a_1$, $b_1$), ($a_2$, $b_2$) into the queue. Once one of the comparisons failed, the overall structural matching failed. This matching process runs iteratively until the queue becomes empty.

The method mentioned in this subsection is called Identical Structure Detection (ISD), which reduces the number of engine calls by one, with the cost of adding two lightweight regularity checking.

\section{Experimental Results for HybridCEC\label{sec:exp}}
This section introduces the experimental settings, followed by the results of the experiments, and some related analyses.

\subsection{Experimental Settings\label{sec:exp-env}}
In this paper, all experiments are carried out on a cluster with two AMD EPYC 7763 CPU @ 2.45Ghz of 128 physical cores in total and 1T RAM under the operating system Ubuntu 20.04 LTS (64bit).

\dpCEC, the CEC solver proposed in this paper, is written in C++ and compiled with GNU g++ (version 9.4.0). It can be accessed from our GitHub repository~\footnote{https://github.com/HybridCEC/Hybrid-CEC}.
The parameter $\rho$ in the selection heuristic is set to 0.15 according to experiments, which will be explained in Section \ref{sec:parameter}.

All the tools and competitors are written in C/C++ and compiled in the same environment as \dpCEC. 

The reasoning tools mentioned in this paper are as follows:
\begin{itemize}
    \item \textbf{\MAB}~\cite{cherif2021kissat}. It is one of the most popular and powerful SAT solvers, which is the winner of the Main-Track of SAT Competition 2021.
    \item \textbf{\minisat}~\cite{een2004extensible}. It can be seen as a milestone of the CDCL SAT solvers, and it is adopted by ABC by default.
    \item \textbf{\kcbox}~\cite{lai2021power}. It is an open-sourced toolbox for knowledge compilation, which integrates some powerful BDD-related tools.
\end{itemize}

We use 50 industrial instances with the application of long bit-wise optimization, which are datapath circuits with adders and multipliers. 
All 50 instances are miter circuits in the AIG format, which are composed of two functional equivalent circuits.
The instances can be divided into 3 classes: $dpm$, $ec$, and $dp$. 
The $dp$ circuits are typical datapath circuits with multiply-add hybrid arithmetic units.
$dpm$ are circuits with a relatively smaller number of arithmetic units, which are mainly multipliers. 
These $dp$ and $dpm$ circuits are generated from an IC company.
The $ec$ circuits can be seen as mixed circuits of $dp$ and $dpm$, and can be downloaded online~\footnote{https://github.com/HybridCEC/Hybrid-CEC/tree/master/ec}.
Note that all the instances from the benchmark are the raw data generated from real-world applications without any simplification or logic synthesis.


For each circuit and a prover/solver, the  runtime limit for each instance is set to 3600 seconds, and `TO' means timeout. \#Solved denotes the number of instances solved by a CEC prover. \#Best denotes the number of instances a solver has the best performance in terms of run time.

\begin{table}[t]
\caption{Comparing \dpCEC with competitors. SAT and BDD are short for the Pure SAT and Pure BDD methods, respectively.
\label{tbl:compare_sota}}
\begin{center}
\setlength{\tabcolsep}{5pt}
\begin{tabular}{|c|c|c|c|c|c|}
\hline
\multirow{2}*{Instance} & \multirow{2}*{Gates} & \multicolumn{4}{|c|}{Solver Name} \\
\cline{3-6} 
 & & \dpCEC &  ABC \&cec & SAT & BDD \\
\hline
dpm\_1\_1 & 386 & \textbf{0.01} & 0.18 & 0.14 & 0.46 \\
dpm\_2\_1 & 867 & \textbf{0.02} & 1.46 & 0.86 & 1.16 \\
dpm\_3\_1 & 696 & \textbf{0.01} & 5.44 & 3.07 & 11.24 \\
dpm\_3\_2 & 975 & \textbf{0.02} & 13.15 & 5.77 & 15.32 \\
dpm\_4\_1 & 877 & \textbf{0.02} & 24.77 & 19.08 & 88.98 \\
dpm\_4\_2 & 1333 & \textbf{0.04} & 60.11 & 21.97 & 81.06 \\
dpm\_4\_3 & 1628 & \textbf{0.08} & 4.88 & 2.82 & 8.67 \\
dpm\_5\_1 & 703 & \textbf{0.01} & 6.08 & 6.02 & 17.33 \\
dpm\_5\_2 & 1319 & \textbf{0.34} & 1576.8 & 834.81 & TO \\
dpm\_5\_3 & 2068 & \textbf{0.84} & 2198.41 & 491.28 & 2207.7 \\
dpm\_6\_1 & 963 & 64.02 & 116.79 & \textbf{57.55} & 252.5 \\
\hline
dp1\_1 & 681 & 82.45 & \textbf{13.93} & 70.67 & 215.29 \\
dp2\_1 & 460 & 1.14 & 2.55 & \textbf{0.59} & 1.19 \\
dp3\_1 & 2116 & \textbf{0.05} & 10.98 & 218.8 & TO \\
dp3\_2 & 2647 & \textbf{0.09} & TO & 538.21 & TO \\
dp3\_3 & 7118 & \textbf{25.7} & TO & TO & TO \\
dp3\_4 & 8574 & \textbf{47.98} & TO & TO & TO \\
dp3\_5 & 10182 & \textbf{42.63} & TO & TO & TO \\
dp4\_1 & 1646 & \textbf{0.05} & 2.52 & 16.93 & 1951.42 \\
dp4\_2 & 5332 & \textbf{24.32} & TO & TO & TO \\
dp4\_3 & 10448 & \textbf{171.28} & TO & TO & TO \\
dp4\_4 & 11256 & \textbf{267.02} & TO & TO & TO \\
dp4\_5 & 12360 & \textbf{487.97} & TO & TO & TO \\
dp5\_1 & 18 & \textbf{$<$0.01} & 0.02 & \textbf{$<$0.01} & 0.18 \\
dp5\_2 & 1646 & \textbf{0.03} & 2.56 & 12.35 & 459.65 \\
dp5\_3 & 9798 & \textbf{424.6} & TO & TO & TO \\
dp5\_4 & 11484 & \textbf{541.12} & TO & TO & TO \\
dp5\_5 & 13617 & \textbf{937.57} & TO & TO & TO \\
dp6\_1 & 4585 & \textbf{2.41} & TO & TO & TO \\
dp6\_2 & 5332 & \textbf{5.85} & TO & TO & TO \\
dp6\_3 & 6128 & \textbf{26.72} & TO & TO & TO \\
dp6\_4 & 8690 & \textbf{297.69} & TO & TO & TO \\
dp6\_5 & 15787 & TO & TO & TO & TO \\
dp7\_1 & 1238 & \textbf{0.03} & 0.36 & 2.41 & 50.3 \\
dp8\_1 & 2116 & \textbf{0.14} & 10.11 & 104.41 & 2532.35 \\
dp9\_1 & 6128 & \textbf{26.78} & TO & TO & TO \\
dp10\_1 & 14049 & TO & TO & TO & TO \\
dp11\_1 & 20091 & TO & TO & TO & TO \\
dp12\_1 & 24773 & TO & TO & TO & TO \\
dp13\_1 & 378 & \textbf{$<$0.01} & 0.02 & 0.02 & 0.3 \\
dp14\_1 & 7061 & \textbf{445.1} & TO & TO & TO \\
\hline
ec\_e1 & 280 & \textbf{$<$0.01} & 0.06 & 0.03 & 0.26 \\
ec\_e2 & 492 & \textbf{0.01} & 0.51 & 0.35 & 0.6 \\
ec\_m1 & 612 & \textbf{$<$0.01} & 0.04 & 0.1 & 0.53 \\
ec\_m2 & 1256 & \textbf{0.02} & 0.41 & 2.17 & 50.37 \\
ec\_m3 & 1664 & \textbf{0.05} & 2.55 & 12.54 & 1314.76 \\
ec\_h1 & 12499 & \textbf{1464.17} & TO & TO & TO \\
ec\_h2 & 13675 & \textbf{3543.39} & TO & TO & TO \\
ec\_h3 & 14152 & TO & TO & TO & TO \\
ec\_h4 & 15604 & \textbf{2497.91} & TO & TO & TO \\
\hline
\multicolumn{2}{|c|}{\#Solved} & \textbf{45} & 25 & 26 & 23\\
\multicolumn{2}{|c|}{\#Best} & \textbf{42} & 1 & 3 & 0\\
\hline
\end{tabular}
\label{tab2}
\end{center}
\end{table}

\begin{table}[t]
\caption{Behavior Statistics of the \dpCEC. Instances that \dpCEC cannot handle are not reported. \label{tbl:info}}
\begin{center}
\begin{tabular}{|c|c|c|c|c|c|c|c|}
\hline
Instance & Gates & Pairs & ISD & \#SAT & $T_{SAT}$ & \#EPS & $T_{EPS}$ \\
 \hline
dpm\_1\_1 & 386 & 3 & 0 & 2 & $<$0.01 & 1 & $<$0.01\\
dpm\_2\_1 & 867 & 6 & 1 & 4 & $<$0.01 & 1 & $<$0.01\\
dpm\_3\_1 & 696 & 5 & 0 & 4 & $<$0.01 & 1 & $<$0.01\\
dpm\_3\_2 & 975 & 6 & 1 & 4 & $<$0.01 & 1 & $<$0.01\\
dpm\_4\_1 & 877 & 6 & 1 & 4 & $<$0.01 & 1 & 0.02\\
dpm\_4\_2 & 1333 & 7 & 1 & 5 & $<$0.01 & 1 & 0.03\\
dpm\_4\_3 & 1628 & 8 & 3 & 4 & $<$0.01 & 1 & 0.05\\
dpm\_5\_1 & 703 & 5 & 0 & 4 & $<$0.01 & 1 & 0.02\\
dpm\_5\_2 & 1319 & 9 & 4 & 4 & $<$0.01 & 1 & 0.34\\
dpm\_5\_3 & 2068 & 13 & 7 & 5 & $<$0.01 & 1 & 0.80\\
dpm\_6\_1 & 963 & 1 & 0 & 1 & 64.23 & 0 & $<$0.01\\
\hline
dp1\_1 & 681 & 1 & 0 & 1 & 82.77 & 0 & $<$0.01\\
dp2\_1 & 460 & 1 & 0 & 1 & 1.13 & 0 & $<$0.01\\
dp3\_1 & 2116 & 40 & 22 & 12 & 0.01 & 6 & $<$0.01\\
dp3\_2 & 2647 & 45 & 25 & 13 & 0.02 & 7 & 0.02\\
dp3\_3 & 7118 & 83 & 47 & 22 & 0.04 & 14 & 25.58\\
dp3\_4 & 8574 & 101 & 59 & 25 & 0.06 & 17 & 47.81\\
dp3\_5 & 10182 & 143 & 83 & 37 & 0.15 & 23 & 42.12\\
dp4\_1 & 1646 & 35 & 19 & 11 & 0.01 & 5 & $<$0.01\\
dp4\_2 & 5332 & 65 & 37 & 17 & 0.02 & 11 & 24.16\\
dp4\_3 & 10448 & 114 & 66 & 29 & 0.06 & 19 & 170.91\\
dp4\_4 & 11256 & 127 & 75 & 31 & 0.21 & 21 & 266.83\\
dp4\_5 & 12360 & 159 & 95 & 39 & 0.25 & 25 & 487.43\\
dp5\_1 & 18 & 1 & 0 & 1 & $<$0.01 & 0 & $<$0.01\\
dp5\_2 & 1646 & 35 & 19 & 11 & 0.01 & 5 & $<$0.01\\
dp5\_3 & 9798 & 98 & 57 & 24 & 0.16 & 17 & 424.23\\
dp5\_4 & 11484 & 119 & 69 & 30 & 0.08 & 20 & 540.56\\
dp5\_5 & 13617 & 171 & 100 & 44 & 43.60 & 27 & 893.61\\
dp6\_1 & 4585 & 60 & 34 & 16 & 0.01 & 10 & 2.23\\
dp6\_2 & 5332 & 65 & 37 & 17 & 0.02 & 11 & 5.73\\
dp6\_3 & 6128 & 70 & 40 & 18 & 0.02 & 12 & 26.60\\
dp6\_4 & 8690 & 85 & 48 & 22 & 0.04 & 15 & 297.50\\
dp7\_1 & 1238 & 30 & 15 & 11 & 0.01 & 4 & $<$0.01\\
dp8\_1 & 2116 & 40 & 22 & 12 & 0.01 & 6 & 0.01\\
dp9\_1 & 6128 & 70 & 40 & 18 & 0.02 & 12 & 26.64\\
dp13\_1 & 378 & 15 & 7 & 6 & $<$0.01 & 2 & $<$0.01\\
dp14\_1 & 7061 & 75 & 43 & 19 & 0.04 & 13 & 445.30\\
\hline
ec\_e1 & 280 & 3 & 0 & 2 & $<$0.01 & 1 & $<$0.01\\
ec\_e2 & 492 & 3 & 0 & 2 & $<$0.01 & 1 & $<$0.01\\
ec\_m1 & 612 & 20 & 10 & 8 & 0.01 & 2 & $<$0.01\\
ec\_m2 & 1256 & 30 & 16 & 10 & $<$0.01 & 4 & $<$0.01\\
ec\_m3 & 1664 & 35 & 18 & 11 & 0.02 & 6 & $<$0.01\\
ec\_h1 & 12499 & 113 & 65 & 28 & 0.06 & 20 & 1465.44\\
ec\_h2 & 13675 & 129 & 76 & 31 & 0.07 & 22 & 3548.04\\
ec\_h4 & 15604 & 163 & 98 & 38 & 0.14 & 27 & 2497.60\\
\hline
\end{tabular}
\label{tab3}
\end{center}
\end{table}

\begin{table}[t]
\caption{Analysis of the Key Techniques used in \dpCEC. \label{tbl:analyse}}
\begin{center}
\setlength{\tabcolsep}{4.0pt}
\begin{tabular}{|c|c|c|c|c|c|c|}
\hline
\multirow{2}*{Instance} & \multirow{2}*{Gates} & \multicolumn{5}{|c|}{Solver Name} \\
\cline{3-7} 
 & & \dpCEC & $V_1$ & $V_2$ & $V_3$ & $V_4$ \\
\hline
dpm\_1\_1 & 386 & \textbf{0.01} & 0.12 & 0.02 & 0.03 & \textbf{0.01} \\
dpm\_2\_1 & 867 & \textbf{0.02} & 0.89 & 0.17 & 0.07 & \textbf{0.02} \\
dpm\_3\_1 & 696 & \textbf{0.01} & 2.82 & 0.08 & 0.04 & \textbf{0.01} \\
dpm\_3\_2 & 975 & 0.02 & 5.55 & 0.13 & 0.07 & \textbf{0.01} \\
dpm\_4\_1 & 877 & \textbf{0.02} & 23.42 & 0.2 & 0.14 & 0.03 \\
dpm\_4\_2 & 1333 & \textbf{0.04} & 27.16 & 0.07 & 0.24 & 0.05 \\
dpm\_4\_3 & 1628 & \textbf{0.08} & 2.9 & 0.12 & 0.28 & 0.09 \\
dpm\_5\_1 & 703 & \textbf{0.01} & 6.43 & 0.06 & 0.11 & 0.03 \\
dpm\_5\_2 & 1319 & \textbf{0.34} & 1334.05 & 5.04 & 3.85 & 0.96 \\
dpm\_5\_3 & 2068 & 0.84 & 681.02 & \textbf{0.31} & 4.84 & 1.09 \\
dpm\_6\_1 & 963 & 64.02 & 64.77 & \textbf{0.79} & 70.36 & 191.45 \\
\hline
dp1\_1 & 681 & \textbf{82.45} & 109.91 & 1469.39 & 89.57 & 307.97 \\
dp2\_1 & 460 & 1.14 & 1.15 & 10.27 & \textbf{1.11} & 2.51 \\
dp3\_1 & 2116 & \textbf{0.05} & 1.98 & TO & 0.22 & 0.13 \\
dp3\_2 & 2647 & \textbf{0.09} & 7.21 & TO & 0.35 & 0.2 \\
dp3\_3 & 7118 & \textbf{25.7} & TO & TO & 120.24 & 64.11 \\
dp3\_4 & 8574 & \textbf{47.98} & TO & TO & 233.99 & 98.52 \\
dp3\_5 & 10182 & \textbf{42.63} & TO & TO & 471.04 & 50.28 \\
dp4\_1 & 1646 & \textbf{0.05} & 0.53 & TO & 0.12 & 0.08 \\
dp4\_2 & 5332 & 24.32 & TO & TO & 103.58 & \textbf{19.81} \\
dp4\_3 & 10448 & 171.28 & TO & TO & 1271.06 & \textbf{85.0} \\
dp4\_4 & 11256 & 267.02 & TO & TO & 2500.96 & \textbf{113.66} \\
dp4\_5 & 12360 & \textbf{487.97} & TO & TO & 3387.28 & 879.55 \\
dp5\_1 & 18 & \textbf{$<$0.01} & \textbf{$<$0.01} & \textbf{$<$0.01} & \textbf{$<$0.01} & \textbf{$<$0.01} \\
dp5\_2 & 1646 & \textbf{0.03} & 0.74 & TO & 0.15 & 0.11 \\
dp5\_3 & 9798 & \textbf{424.6} & TO & TO & 2813.27 & 1104.36 \\
dp5\_4 & 11484 & \textbf{541.12} & TO & TO & TO & 1535.88 \\
dp5\_5 & 13617 & \textbf{937.57} & TO & TO & TO & 2722.96 \\
dp6\_1 & 4585 & \textbf{2.41} & 1220.27 & TO & 27.9 & 3.14 \\
dp6\_2 & 5332 & \textbf{5.85} & TO & TO & 169.79 & 21.32 \\
dp6\_3 & 6128 & \textbf{26.72} & TO & TO & 588.2 & 232.32 \\
dp6\_4 & 8690 & \textbf{297.69} & TO & TO & TO & 1765.74 \\
dp6\_5 & 15787 & TO & TO & TO & TO & TO \\
dp7\_1 & 1238 & \textbf{0.03} & 0.2 & TO & 0.09 & 0.08 \\
dp8\_1 & 2116 & \textbf{0.14} & 2.0 & TO & 0.47 & 0.16 \\
dp9\_1 & 6128 & 26.78 & TO & TO & 889.63 & \textbf{24.9} \\
dp10\_1 & 14049 & TO & TO & TO & TO & TO \\
dp11\_1 & 20091 & TO & TO & TO & TO & TO \\
dp12\_1 & 24773 & TO & TO & TO & TO & TO \\
dp13\_1 & 378 & \textbf{$<$0.01} & 0.02 & 0.42 & 0.01 & 0.02 \\
dp14\_1 & 7061 & 445.1 & TO & TO & 3330.51 & \textbf{157.6} \\
\hline
ec\_e1 & 280 & \textbf{$<$0.01} & 0.04 & 0.02 & 0.03 & 0.01 \\
ec\_e2 & 492 & \textbf{0.01} & 0.28 & 0.06 & 0.03 & \textbf{0.01} \\
ec\_m1 & 612 & \textbf{$<$0.01} & 0.02 & 29.89 & 0.05 & 0.01 \\
ec\_m2 & 1256 & \textbf{0.02} & 0.1 & TO & 0.1 & 0.05 \\
ec\_m3 & 1664 & \textbf{0.05} & 0.44 & TO & 0.19 & \textbf{0.05} \\
ec\_h1 & 12499 & 1464.17 & TO & TO & TO & \textbf{1061.78} \\
ec\_h2 & 13675 & \textbf{3543.39} & TO & TO & TO & TO \\
ec\_h3 & 14152 & TO & TO & TO & TO & TO \\
ec\_h4 & 15604 & \textbf{2497.91} & TO & TO & TO & 3161.23 \\
\hline
\multicolumn{2}{|c|}{\#Solved} & \textbf{45}     & 27       & 18 & 39 & 44 \\
\multicolumn{2}{|c|}{\#Best}   & \textbf{35} & 1 & 3 & 2 & 13\\
\hline
\end{tabular}
\label{tab4}
\end{center}
\end{table}

\subsection{Evaluation on Datapath Benchmarks}

We compare our solver, \dpCEC, with other state-of-the-art solvers, which are as follows:
\begin{itemize}
    \item \textbf{ABC \&cec}~\cite{mishchenko2006improvements}. It is the main competitor in this paper, which is one of the most representative SAT sweeping based CEC algorithms.
    \item \textbf{Pure SAT}. Using an effective SAT solver, \MAB~\cite{cherif2021kissat}, to check $M_o$ directly without sweeping.
    \item \textbf{Pure BDD}. Using an effective BDD solver, \kcbox~\cite{lai2021power}, to check $M_o$ directly without sweeping.
\end{itemize}

From the results in Table~\ref{tbl:compare_sota}, we learn that \dpCEC significantly outperforms other competitors. On the 50 instances, \dpCEC solves 20, 19, and 22 more instances than ABC \&cec, pure SAT, and pure BDD, separately.
Compared with ABC \&cec, there are 12\% of the instances on which \dpCEC is at least 3 orders of magnitude faster, while there are 30\% of the instances on which \dpCEC is at least 2 orders of magnitude faster. Besides, it shows that the other methods have similar performance on this datapath circuits benchmark.

\subsection{Behavior Statistics Analysis}

For better learning of the behavior in the process of \dpCEC, we collect the statistics during the running process of \dpCEC. In Table~\ref{tbl:info}, we report the number of potential-equivalent node pairs (`Pairs' column), the number of pairs reduced by ISD (`ISD' column), the number of SAT solver calls (`\#SAT' column), the time used in SAT solving in second (`$T_{SAT}$' column), the number of EPS calls (`\#EPS' column), and the Accumulated running time (`$T_{EPS}$' column).

From the `ISD' column, we learn that a large proportion ($56.4\%$) of the sub-circuits in the datapath circuits share the same implementation with other sub-circuits. With ISD, a significant amount of time can be saved by reducing unnecessary engine calls.

In the datapath circuits, the engine selection heuristic chooses EPS when there is a high density of XOR chains. The proportion of the EPS calls is a reflection of the difficulty of a circuit. From the table, we can learn that EPS plays an important role in the verification process. Meanwhile, we can learn that the verification difficulty for a datapath circuit is mainly due to the fact that the overall verification process is always stuck at some hard-to-prove sub-circuits.

\subsection{Strategy Assessments}

This subsection evaluates the effectiveness of each strategy used in \dpCEC. In Table~\ref{tbl:analyse}, we compare \dpCEC with its 4 variants:
\begin{itemize}
    \item $V_1$: using only SAT solver in sweeping.
    \item $V_2$: using only EPS to prove in sweeping.
    \item $V_3$: disable the ISD technique.
    \item $V_4$: replacing the SAT solver with \minisat~\cite{een2004extensible}.
\end{itemize}

By comparing \dpCEC with $V_1$ and $V_2$, we learn that the performance of \dpCEC has been significantly improved thanks to the selection heuristic.
From the comparison between \dpCEC and $V_3$, we learn that ISD improves the speed of the verification process.
For fairness of the comparison with ABC, which integrates \minisat as its SAT solver, we change our SAT solver from \MAB~\cite{cherif2021kissat} to \minisat. From the result of $V_4$, we learn that
extremely high SAT solving performance is not crucial for \dpCEC on those datapath circuits,
which proves the effectiveness of our framework and internal strategies.

\subsection{Analysis for Selection Heuristic\label{sec:parameter}}
In this section, we demonstrate more details about the parameter setting of the selection heuristic.
Section~\ref{sec:motivation} shows the complementary abilities between SAT and EPS on different types of circuits, and Section~\ref{sec:select} proposed the $score_{XOR}$ based on XOR chain density.

\begin{figure}[t]
\centerline{\includegraphics[width=0.35\textwidth]{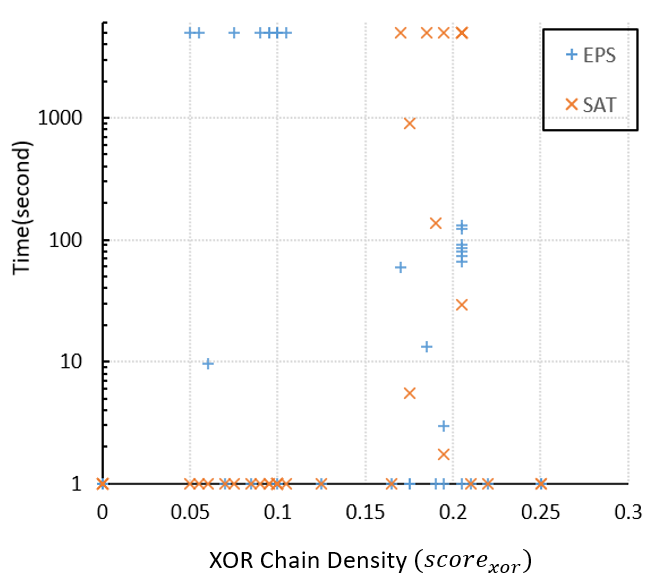}}
\caption{The runtime comparison between EPS and SAT for each potential-equivalent internal node pair, where the nodes are sorted according to $score_{xor}$.}
\label{fig:xor-score}
\end{figure}

Figure~\ref{fig:xor-score} is the scatter plot that presents the relationship between the runtime and the $score_{XOR}$. For each potential-equivalent pair of the instance `ec\_h1', we calculate the related $score_{XOR}$ and the runtime for SAT and EPS, and draw the points on the figure. From the results, we learn that SAT solvers are more suitable for circuits with $score_{XOR}\leq0.15$, while EPS is more suitable for circuits with $score_{XOR}>0.15$. We note that the datapath circuits in this paper share a similar phenomenon as `ec\_h1'.

\section{Parallel Extension for Hybrid CEC Algorithm\label{sec:parahcec}}
On the basis of \dpCEC, we parallelized the internal SAT and EPS engines, resulting in the parallel CEC prover \parCEC. 

\begin{figure}[t]
\centering
\centerline{\includegraphics[width=0.25\textwidth]{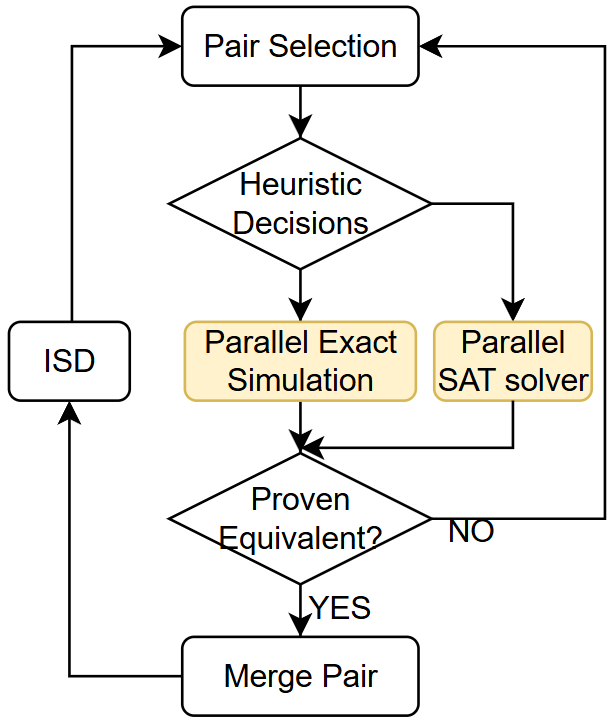}}
\caption{Main Framework for Our Parallel CEC Algorithm.}
\label{fig:framework3}
\end{figure}

The framework of \parCEC is presented in Figure \ref{fig:framework3}, the differences between \dpCEC and \parCEC are as follows.
\begin{itemize}
    \item Replace the internal SAT solver from a sequential SAT solver to a high-performance parallel SAT solver called PRS~\cite{chen2023prs}.
    \item Parallelize the internal EPS engine by grouping the PI patterns and performing simulation separately.
\end{itemize}

\subsection{Parallel SAT Engine}
For each internal potential equivalent node that is checked with the SAT engine, we replace the sequential SAT solver with the state-of-the-art parallel SAT solver PRS~\cite{chen2023prs}.

PRS~\cite{chen2023prs} is a lightweight, generic, and powerful parallel SAT framework based on Preprocessing and Regular Shifting, which has won the SAT competition in 2022 and 2023 by a considerable margin.

\begin{figure}[t]
\centering
\centerline{\includegraphics[width=0.5\textwidth]{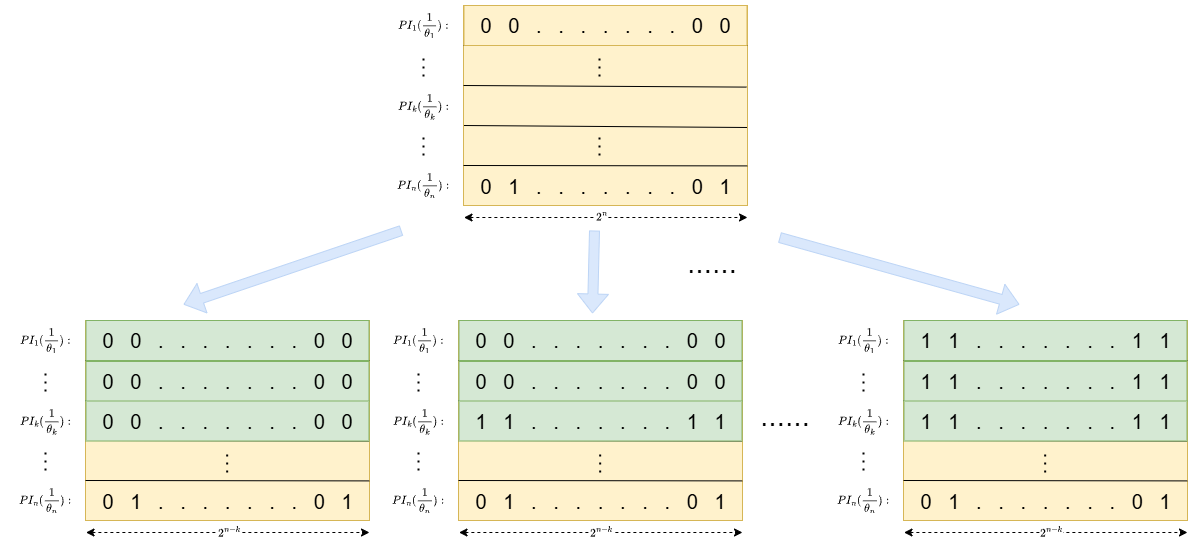}}
\caption{Main Framework for Our Parallel CEC Algorithm.}
\label{fig:eps-split}
\end{figure}

\subsection{Parallelize EPS Engine}
For EPS, We enumerated the probability signals of the first $k$ PIs to divide the overall task into $2^k$ subtasks, note that this partitioning method is naturally lossless and load-balanced. In detail, as seen from Figure \ref{fig:eps-split}, suppose there are $2^k$ available threads, then we fixed the value of $PI_0,PI_1,...PI_{k-1}$ of these threads as $(PI_0=0,PI_1=0,\ldots,PI_{k-1}=0),(PI_0=0,PI_1=0,\ldots,PI_{k-1}=1),\ldots,(PI_0=1,PI_1=1,\ldots,PI_{k-1}=1)$. Therefore the difficulty of each subtask solved by each thread will be reduced to $1/{2^k}$ of the overall task.

\section{Experimental Results for \parCEC\label{sec:paraexp}}
In this section,  We evaluate the performance of \parCEC and compare it with \dpCEC and ABC \&splitprove, the parallel version of ABC \&cec~\cite{mishchenko2006improvements}. In addition, We perform experimental analyses on the strategies in \parCEC. For the parallel algorithm, each thread occupied a single core. The experimental environment, experimental settings and benchmarks remained consistent with the previous setup.

\subsection{Evaluation on Datapath Benchmarks}
\begin{table*}[t]
\caption{Comparing \parCEC with parallel version of ABC \&cec and \dpCEC.}
\label{tbl:compare parCEC}
\begin{center}
\setlength{\tabcolsep}{5pt}
\begin{tabular}{|c|c|c|c|c|c|c|c|c|c|c|}
\hline
\multirow{2}*{Instance} & \multirow{2}*{Gates} & \multicolumn{2}{c|}{ABC \&splitprove} & \multirow{2}*{\dpCEC} & \multicolumn{6}{c|}{\parCEC} \\
\cline{3-4}\cline{6-11} 
 & & 32t & 64t & & 2t & 4t & 8t & 16t & 32t & 64t\\
\hline
dpm\_1\_1 & 386 & 0.16 & 0.16 & \textbf{0.01} & 0.23 & 0.14 & 0.04 & 0.04 & 0.04 & 0.04 \\
dpm\_2\_1 & 867 & 1.1 & 1.19 & \textbf{0.02} & 0.46 & 0.16 & 0.07 & 0.06 & 0.07 & 0.08 \\
dpm\_3\_1 & 696 & 3.56 & 3.83 & \textbf{0.01} & 0.45 & 0.25 & 0.08 & 0.05 & 0.06 & 0.07 \\
dpm\_3\_2 & 975 & 9.43 & 25.8 & \textbf{0.02} & 0.47 & 0.16 & 0.07 & 0.07 & 0.07 & 0.09 \\
dpm\_4\_1 & 877 & 48.26 & 47.85 & \textbf{0.02} & 0.47 & 0.36 & 0.08 & 0.07 & 0.07 & 0.08 \\
dpm\_4\_2 & 1333 & 43.95 & 43.9 & \textbf{0.04} & 0.59 & 0.29 & 0.1 & 0.1 & 0.1 & 0.11 \\
dpm\_4\_3 & 1628 & 3.86 & 4.17 & \textbf{0.08} & 0.5 & 0.3 & 0.11 & 0.11 & 0.1 & 0.13 \\
dpm\_5\_1 & 703 & 29.3 & 28.61 & \textbf{0.01} & 0.35 & 0.35 & 0.07 & 0.06 & 0.06 & 0.07 \\
dpm\_5\_2 & 1319 & 120.79 & 96.8 & 0.34 & 0.61 & 0.35 & 0.15 & \textbf{0.11} & 0.11 & 0.12 \\
dpm\_5\_3 & 2068 & 105.53 & 94.84 & 0.84 & 0.83 & 0.54 & 0.21 & 0.17 & \textbf{0.16$^{*}$} & 0.18 \\
dpm\_6\_1 & 963 & 59.85 & 58.97 & 64.02 & 40.01 & 27.59 & 18.01 & 10.4$^{*}$ & 6.81$^{*}$ & \textbf{5.64$^\dag$} \\
\hline
dp1\_1 & 681 & 114.98 & 114.49 & 82.45 & 60.64 & 49.39 & 36.17 & 26.5 & 14.2$^{*}$ & \textbf{11.62$^{*}$} \\
dp2\_1 & 460 & 1.31 & 0.95 & 1.14 & 0.73 & 0.74 & 0.74 & 0.76 & 0.75 & \textbf{0.67} \\
dp3\_1 & 2116 & 1334.38 & 619.12 & \textbf{0.05} & 1.32 & 1.13 & 0.54 & 0.47 & 0.36 & 0.41 \\
dp3\_2 & 2647 & TO & TO & \textbf{0.09} & 1.45 & 1.45 & 0.56 & 0.51 & 0.29 & 0.45 \\
dp3\_3 & 7118 & TO & TO & 25.7 & 6.73 & 4.39$^{*}$ & 2.18$^\dag$ & 1.49$^\dag$ & \textbf{0.98$^\ddag$} & 1.03$^\ddag$ \\
dp3\_4 & 8574 & TO & TO & 47.98 & 9.99 & 6.5$^{*}$ & 3.54$^\dag$ & 2.03$^\ddag$ & \textbf{1.31$^\ddag$} & 1.39$^\ddag$ \\
dp3\_5 & 10182 & TO & TO & 42.63 & 19.31 & 11.61 & 6.6$^{*}$ & 3.62$^\dag$ & 2.38$^\dag$ & \textbf{2.3$^\dag$} \\
dp4\_1 & 1646 & 118.79 & 118.2 & \textbf{0.05} & 1.2 & 1.2 & 0.41 & 0.45 & 0.25 & 0.4 \\
dp4\_2 & 5332 & TO & TO & 24.32 & 5.21 & 3.61$^{*}$ & 1.78$^\dag$ & 1.17$^\ddag$ & \textbf{0.79$^\ddag$} & 0.86$^\ddag$ \\
dp4\_3 & 10448 & TO & TO & 171.28 & 42.71 & 24.62$^{*}$ & 12.81$^\dag$ & 6.93$^\ddag$ & 3.79$^\ddag$ & \textbf{2.75$^\ddag$} \\
dp4\_4 & 11256 & TO & TO & 267.02 & 55.46 & 32.05$^{*}$ & 16.41$^\dag$ & 9.07$^\ddag$ & 4.54$^\ddag$ & \textbf{3.61$^\ddag$} \\
dp4\_5 & 12360 & TO & TO & 487.97 & 81.64$^{*}$ & 46.51$^\dag$ & 24.37$^\ddag$ & 13.65$^\ddag$ & 6.64$^\ddag$ & \textbf{5.16$^\ddag$} \\
dp5\_1 & 18 & 0.04 & 0.35 & \textbf{$<$0.01} & 0.11 & 0.12 & 0.01 & 0.02 & 0.02 & 0.03 \\
dp5\_2 & 1646 & 118.25 & 117.43 & \textbf{0.03} & 1.19 & 0.9 & 0.41 & 0.38 & 0.35 & 0.4 \\
dp5\_3 & 9798 & TO & TO & 424.6 & 108.9 & 60.57$^{*}$ & 31.32$^\dag$ & 17.35$^\ddag$ & 8.21$^\ddag$ & \textbf{4.94$^\ddag$} \\
dp5\_4 & 11484 & TO & TO & 541.12 & 182.84 & 100.74$^{*}$ & 51.9$^\dag$ & 28.23$^\dag$ & 13.13$^\ddag$ & \textbf{7.84$^\ddag$} \\
dp5\_5 & 13617 & TO & TO & 937.57 & 411.89 & 237.88 & 136.7$^{*}$ & 73.98$^\dag$ & 38.0$^\ddag$ & \textbf{24.73$^\ddag$} \\
dp6\_1 & 4585 & TO & TO & 2.41 & 2.48 & 1.89 & 1.03 & 0.73 & 0.67 & \textbf{0.66} \\
dp6\_2 & 5332 & TO & TO & 5.85 & 5.16 & 3.74 & 1.86 & 1.18 & 0.8$^{*}$ & \textbf{0.78$^{*}$} \\
dp6\_3 & 6128 & TO & TO & 26.72 & 17.3 & 10.36 & 5.49 & 2.86$^{*}$ & 1.63$^\dag$ & \textbf{1.34$^\dag$} \\
dp6\_4 & 8690 & TO & TO & 297.69 & 245.09 & 137.56 & 73.3 & 40.61$^{*}$ & 20.41$^\dag$ & \textbf{11.68$^\ddag$} \\
dp6\_5 & 15787 & TO & TO & TO & 1941.15$^\ddag$ & 1097.49$^\ddag$ & 572.23$^\ddag$ & 293.87$^\ddag$ & 151.56$^\ddag$ & \textbf{76.01$^\ddag$} \\
dp7\_1 & 1238 & 7.74 & 7.3 & \textbf{0.03} & 1.17 & 0.98 & 0.49 & 0.42 & 0.42 & 0.23 \\
dp8\_1 & 2116 & 1426.28 & 659.65 & \textbf{0.14} & 1.31 & 1.22 & 0.43 & 0.48 & 0.36 & 0.39 \\
dp9\_1 & 6128 & TO & TO & 26.78 & 17.44 & 10.37 & 5.37 & 3.02$^{*}$ & 1.76$^\dag$ & \textbf{1.29$^\ddag$} \\
dp13\_1 & 378 & 0.07 & 0.06 & \textbf{$<$0.01} & 0.63 & 0.54 & 0.24 & 0.25 & 0.05 & 0.07 \\
dp14\_1 & 7061 & TO & TO & 445.1 & 72.25$^{*}$ & 40.63$^\dag$ & 21.36$^\ddag$ & 11.78$^\ddag$ & 5.67$^\ddag$ & \textbf{3.65$^\ddag$} \\
\hline
ec\_e1 & 280 & 0.06 & 0.06 & \textbf{$<$0.01} & 0.23 & 0.13 & 0.03 & 0.03 & 0.04 & 0.04 \\
ec\_e2 & 492 & 0.27 & 0.25 & \textbf{0.01} & 0.24 & 0.14 & 0.04 & 0.04 & 0.04 & 0.05 \\
ec\_m1 & 612 & 0.23 & 0.22 & \textbf{$<$0.01} & 0.84 & 0.75 & 0.46 & 0.36 & 0.17 & 0.18 \\
ec\_m2 & 1256 & 5.88 & 5.58 & \textbf{0.02} & 1.08 & 1.08 & 0.4 & 0.4 & 0.31 & 0.34 \\
ec\_m3 & 1664 & 119.78 & 119.02 & \textbf{0.05} & 1.2 & 1.1 & 0.43 & 0.42 & 0.34 & 0.28 \\
ec\_h1 & 12499 & TO & TO & 1464.17 & 702.64 & 397.54 & 202.8$^{*}$ & 105.9$^\dag$ & 49.05$^\ddag$ & \textbf{26.59$^\ddag$} \\
ec\_h2 & 13675 & TO & TO & 3543.39 & 977.86 & 531.71$^{*}$ & 269.01$^\dag$ & 142.42$^\ddag$ & 65.15$^\ddag$ & \textbf{34.77$^\ddag$} \\
ec\_h3 & 14152 & TO & TO & TO & 1071.77$^\ddag$ & 605.86$^\ddag$ & 304.65$^\ddag$ & 162.15$^\ddag$ & 75.95$^\ddag$ & \textbf{40.74$^\ddag$} \\
ec\_h4 & 15604 & TO & TO & 2497.91 & 1634.39 & 860.98 & 460.11$^{*}$ & 245.86$^\dag$ & 120.0$^\ddag$ & \textbf{59.79$^\ddag$} \\
\hline
\multicolumn{2}{|c|}{\#Solved} & 25 & 25 & 45 & \textbf{47} & \textbf{47} & \textbf{47} & \textbf{47} & \textbf{47} & \textbf{47}\\
\multicolumn{2}{|c|}{PAR2} & 3448.38 & 3416.36 & 549.57 & 164.5 & 91.9 & 48.2 & 25.8 & 12.7 & \textbf{7.1} \\
\hline
\multicolumn{9}{l}{* 5x faster than \dpCEC} \\
\multicolumn{9}{l}{$\dag$ 10x faster than \dpCEC} \\
\multicolumn{9}{l}{$\ddag$ 20x faster than \dpCEC} \\
\end{tabular}
\label{tab2}
\end{center}
\end{table*}

We compared different thread versions of \parCEC with ABC \&splitprove and \dpCEC. In the subsequent tables, we omitted the results for instances dp10\_1, dp11\_1 and dp12\_1 because no solver was able to solve them within the given runtime limit. We also report the penalized average run time score (PAR2), which penalizes the run time of a failed run as twice the runtime limit. 

From Table \ref{tbl:compare parCEC}, we can see that the PAR2 decreases as the number of threads increases. Specifically, the speedup of \parCEC with 2 (4, 8, 16, 32, 64) threads is 3.3 (6.0, 11.4, 21.3, 43.3, 77.4) according to the PAR2. Furthermore, \parCEC performs significantly better than ABC \&splitprove. These results verify the effectiveness and scalability of \parCEC.

\subsection{Strategy Assessments}
To evaluate the effectiveness of the parallelization strategy in \parCEC. We compare the 32 threads version of \parCEC with its 2 variants:
\begin{itemize}
    \item $V_1$: replacing the parallel SAT solver with sequential SAT solver \textbf{\MAB}~\cite{cherif2021kissat}.
    \item $V_2$: replacing the parallel EPS engine with sequential EPS engine.
\end{itemize}

The comparison of \parCEC and its variants is presented in Table \ref{tbl:paraanalyse}, the performance of each variant is far weaker than \parCEC, which demonstrates the effectiveness of the parallelized strategies especially the parallelized EPS engine.

\begin{table}[t]
\caption{Analysis of the parallelized strategies used in \parCEC. \label{tbl:paraanalyse}}
\begin{center}
\setlength{\tabcolsep}{4.0pt}
\begin{tabular}{|c|c|c|c|c|}
\hline
\multirow{2}*{Instance} & \multirow{2}*{Gates} & \multicolumn{3}{c|}{Solver Name} \\
\cline{3-5} 
 & & \parCEC(32t) & $V_1$(32t) & $V_2$(32t) \\
\hline
dpm\_1\_1 & 386 & \textbf{0.04} & 0.08 & 0.06 \\
dpm\_2\_1 & 867 & \textbf{0.07} & 1.15 & 0.08 \\
dpm\_3\_1 & 696 & 0.06 & 1.14 & \textbf{0.05} \\
dpm\_3\_2 & 975 & 0.07 & 1.16 & \textbf{0.07} \\
dpm\_4\_1 & 877 & 0.07 & 1.17 & \textbf{0.06} \\
dpm\_4\_2 & 1333 & 0.1 & 1.23 & \textbf{0.09} \\
dpm\_4\_3 & 1628 & \textbf{0.1} & 1.22 & 0.11 \\
dpm\_5\_1 & 703 & 0.06 & 1.15 & \textbf{0.05} \\
dpm\_5\_2 & 1319 & 0.11 & 1.66 & \textbf{0.1} \\
dpm\_5\_3 & 2068 & 0.16 & 2.19 & \textbf{0.15} \\
dpm\_6\_1 & 963 & \textbf{6.81} & 8.32 & 57.83 \\
\hline
dp1\_1 & 681 & \textbf{14.2} & 24.14 & 71.33 \\
dp2\_1 & 460 & \textbf{0.75} & 1.07 & 1.11 \\
dp3\_1 & 2116 & 0.36 & 6.41 & \textbf{0.14} \\
dp3\_2 & 2647 & 0.29 & 6.96 & \textbf{0.17} \\
dp3\_3 & 7118 & 0.98 & 31.39 & \textbf{0.69} \\
dp3\_4 & 8574 & 1.31 & 50.26 & \textbf{1.1} \\
dp3\_5 & 10182 & 2.38 & 95.69 & \textbf{1.81} \\
dp4\_1 & 1646 & 0.25 & 5.85 & \textbf{0.12} \\
dp4\_2 & 5332 & 0.79 & 26.26 & \textbf{0.56} \\
dp4\_3 & 10448 & 3.79 & 224.1 & \textbf{3.39} \\
dp4\_4 & 11256 & 4.54 & 291.82 & \textbf{4.37} \\
dp4\_5 & 12360 & 6.64 & 438.76 & \textbf{6.4} \\
dp5\_1 & 18 & \textbf{0.02} & 0.54 & 0.02 \\
dp5\_2 & 1646 & 0.35 & 5.85 & \textbf{0.11} \\
dp5\_3 & 9798 & 8.21 & 582.66 & \textbf{7.98} \\
dp5\_4 & 11484 & \textbf{13.13} & 972.11 & 14.0 \\
dp5\_5 & 13617 & \textbf{38.0} & 2051.29 & 69.69 \\
dp6\_1 & 4585 & 0.67 & 11.94 & \textbf{0.32} \\
dp6\_2 & 5332 & 0.8 & 26.81 & \textbf{0.55} \\
dp6\_3 & 6128 & 1.63 & 88.55 & \textbf{1.44} \\
dp6\_4 & 8690 & \textbf{20.41} & 1198.34 & 20.43 \\
dp6\_5 & 15787 & \textbf{151.56} & TO & 189.53 \\
dp7\_1 & 1238 & 0.42 & 5.83 & \textbf{0.09} \\
dp8\_1 & 2116 & 0.36 & 6.4 & \textbf{0.13} \\
dp9\_1 & 6128 & 1.76 & 86.79 & \textbf{1.51} \\
dp13\_1 & 378 & 0.05 & 3.19 & \textbf{0.04} \\
dp14\_1 & 7061 & 5.67 & 393.4 & \textbf{5.38} \\
\hline
ec\_e1 & 280 & \textbf{0.04} & 0.07 & \textbf{0.04} \\
ec\_e2 & 492 & \textbf{0.04} & 0.08 & \textbf{0.04} \\
ec\_m1 & 612 & 0.17 & 4.25 & \textbf{0.06} \\
ec\_m2 & 1256 & 0.31 & 5.32 & \textbf{0.09} \\
ec\_m3 & 1664 & 0.34 & 5.85 & \textbf{0.11} \\
ec\_h1 & 12499 & \textbf{49.05} & TO & 50.35 \\
ec\_h2 & 13675 & \textbf{65.15} & TO & 67.22 \\
ec\_h3 & 14152 & \textbf{75.95} & TO & 76.61 \\
ec\_h4 & 15604 & 120.0 & TO & \textbf{114.62} \\
\hline
\multicolumn{2}{|c|}{\#Solved} & \textbf{47}     & 42 & \textbf{47} \\
\multicolumn{2}{|c|}{PAR2}  & \textbf{12.72} & 907.92 & 16.39\\
\hline
\end{tabular}
\label{tab4}
\end{center}
\end{table}

\section{Conclusions\label{sec:concl}}
In this paper, we implemented a CEC prover, \dpCEC, which improves the popular SAT Sweeping based framework by integrating an exact probability-based simulation (EPS) method. Previous EPS could only handle circuits within 24 PIs due to memory limits, thus we optimize the memory usage by truncating the input values, making EPS suitable for circuits with any number of PIs. Then, for better cooperation between EPS and SAT in the sweeping process, a selection heuristic is designed based on XOR chain density. Moreover, \dpCEC is improved according to the regularity in datapath circuits. Extensive experiments on industrial datapath circuits are carried out to prove the effectiveness of \dpCEC. Besides, on top of \dpCEC, we developed the parallel CEC prover, \parCEC, which parallelized the internal SAT and EPS engines. Experimental results demonstrate the efficiency and stability of \parCEC.

In the future, we plan to develop a distributed version of \parCEC and attempt to migrate the techniques proposed in this paper into the sequential equivalence checking problem.

\section{Acknowledgment}
This work was supported by the Strategic Priority Research Program of the Chinese Academy of Sciences, Grant No. XDA0320000 and XDA0320300. We thank the anonymous reviewers for their constructive comments and suggestions.

\bibliographystyle{IEEEtran}
\bibliography{cec}

\end{document}